\documentclass[reprint,twocolumn,longbibliography,floatfix,amsmath,amssymb, aps,pra]{revtex4-1}

\usepackage{soul}
\usepackage{amsmath}
\usepackage{amssymb}
\usepackage{mathtools}
\usepackage{graphicx}
\usepackage{dcolumn}
\usepackage{bm}
\usepackage{amsmath}
\usepackage{graphicx}
\usepackage{amsfonts}
\usepackage{subfigure}
\usepackage{graphicx}
\usepackage{array}
\usepackage{float}
\usepackage{color}
\usepackage{amssymb}%
\usepackage[colorlinks=true,linkcolor=blue]{hyperref}%
\hypersetup{allcolors=blue}
\usepackage[normalem]{ulem}
\usepackage{xcolor}

\usepackage{mathtools}
\DeclarePairedDelimiterX\braket[2]{\langle}{\rangle}{#1 \delimsize\vert #2}

\begin{document}
\preprint{APS/123-QED}
\title{Rydberg-EIT of $^{85}$Rb vapor in a cell with Ne buffer gas}
\date{\today }

\author{Nithiwadee Thaicharoen$^{1}$}
    \email{nithi@umich.edu}
\affiliation{Department of Physics and Materials Science, Faculty of Science, Chiang Mai University, Chiang Mai 50200, Thailand}
\author{Ryan Cardman$^{2}$}
\author{Georg Raithel$^{2}$}
\email{graithel@umich.edu}
\affiliation{Department of Physics, University of Michigan, Ann Arbor, MI 48109, USA}

\begin{abstract}
We investigate Rydberg electromagnetically induced transparency (EIT) of $^{85}$Rb atomic vapor in a glass cell that contains a 5-Torr neon buffer gas. At low probe power,
EIT lines exhibit a positive frequency shift of about 70~MHz and a broadening of about 120~MHz, with minimal dependence on the principal quantum number of the Rydberg states. The EIT line shift arises from s-wave scattering between the Rydberg electron and the Ne atoms, which induces a positive shift near 190~MHz, and from the polarization of the Ne atoms within the Rydberg atom, which adds a negative shift near -120~MHz. The line broadening is largely due to the Ne polarization. Our experimental results are in good qualitative agreement with our theoretical model, in which the shift is linear in buffer-gas density. Our results suggest that Rydberg-EIT can serve as a direct spectroscopic probe for buffer-gas density at low pressure, and that it is suitable for non-invasive measurement of electric fields in low-pressure noble-gas discharge plasmas and in dusty plasmas.

\end{abstract}
\maketitle


Electromagnetically induced transparency (EIT) involving Rydberg-atom spectroscopy in room-temperature atomic vapors has become an important technique for electric-field sensing, offering sensitivity in metrology applications~\cite{sedlacek12, sedlacek13, holloway14} as well as avenues towards non-traditional radio reception~\cite{Anderson2020, Meyer2021, receiver2021}. 
Rydberg Stark spectroscopy and direct-particle detection was used to measure macroscopic and microscopic electric fields in ion plasmas prepared from laser-cooled atom clouds~\cite{Feldbaum2002, Duspayev2023}. In plasmas generated in thermal atomic vapors, Rydberg-EIT Stark spectroscopy was employed for non-invasive all-optical electric-field measurement~\cite{anderson17, weller19}. DC electric fields of charges released by photo-illumination of a borosilicate vapor cell were analyzed using Rydberg-EIT spectroscopy of Rb $nD_{J}$ Rydberg levels~\cite{ma2020dc}. These developments suggest that Rydberg-EIT has the potential to serve as a non-invasive plasma electric-field probe in glass tubes, vacuum systems, or vapor cells that harbor a low-pressure discharge or an inductively coupled plasma, including RF or DC rubidium plasma lamps~\cite{bell1961alkali, brewer1961high} that are commonly used as spectroscopic frequency references and for optical pumping of alkali vapors in magnetic-field-sensing cells. Moreover, plasma often contains charged dust particles~\cite{Shukla2009}. Such dusty plasmas appear, for instance, in astrophysical settings (including the B ring of Saturn~\cite{mitchell2006saturn}, Martian dust devils~\cite{renno2003electrical} and the moon~\cite{Popel2018}), as well as in technical plasmas (including fusion~\cite{Ratynskaia2022} and microfabrication devices~\cite{Boufendi2011}). The dynamics of such plasmas are being explored in ground-based~\cite{merlino2009dust, menati2019filamentation} and microgravity setups~\cite{thoma2023complex} using low-pressure noble-gas plasmas, which are seeded with dust particles. Rydberg-EIT could be ideal for non-intrusive measurement of electric fields, Debye shielding and particle interaction~\cite{Sheridan2019}, and electric-field wakes~\cite{Joshi2023} in a dusty plasma.

As a step towards Rydberg-EIT field sensing in these systems, it is required to assess the viability of Rydberg-EIT of a suitable atomic-sensor species, such as rubidium or cesium, in a noble-gas background with pressures ranging from tens of milli-Torr to several Torr. This necessitates a study of the effects of the background buffer gas on the Rydberg-EIT spectrum. Previous work~\cite{Sargsyan2010} has explored the effects of Rb-Ne collisions on $5S_{1/2}\rightarrow5P_{3/2}\rightarrow5D_{5/2}$ EIT linewidths in a buffer-gas cell. In our present work, we demonstrate a first observation of $^{85}$Rb Rydberg-EIT at principal quantum numbers of $n \sim 40$ in a cell with a 5-Torr neon buffer gas. We measure the frequency shift and line broadening of the EIT signal due to the background gas. At high probe-laser power, we observe a transition from EIT to electromagnetically induced absorption (EIA). The observed features do not significantly depend $n$. Our studies present a stepping stone towards employing Rydberg-EIT as a versatile tool for characterizing electric fields in dusty plasma, which can be prepared even at noble-gas pressures substantially below 5~Torr. 

It is noted that vapor cells consisting of an alkali metal mixed with a higher-pressure inert buffer gas are of interest in other applications that require reduced ground-state spin relaxation rates, such as Faraday~\cite{Budker2002}  and SERF~\cite{Allred2002} magnetometers. Optical pumping of the alkali vapor can also be used to spin-polarize the noble gas via spin-exchange collisions, a method conducive to NMR with optically prepared spin-polarized gases~\cite{Kornack2005}.

\section{Theory}
\label{sec:theory}
Atoms in highly-excited Rydberg states exhibit sensitivity to their  environment. When the Rydberg atoms are perturbed by dense ground-state perturbers, their interaction with the surrounding medium gives rise to a frequency shift, which can be attributed to two main effects~\cite{alekseev1966spectroscopic,Omont1977,Asaf1993}. The dominant effect arises from the scattering of the Rydberg electron by the perturbers within a Rydberg-atom volume of $\sim \frac{4}{3}\pi (2n^2 a_0)^3$. This scattering effect can be explained by a Fermi interaction~\cite{fermi1934sopra}, and the resulting angular frequency shift in units of rad/s is given by

\begin{equation}
\Delta\omega_{\text{sc}} =
2 \pi a_{s} N
\left[\frac{e^{2} a_0}{ 4\pi\epsilon_0 \hbar }\right] \quad ,
\label{eq:delta_sc}
\end{equation}
where $\hbar$ is the reduced Planck's constant,
$e$ is the elementary charge in C,
$a_{s}$ is the low-energy s-wave scattering length in meters, 
and $N$ is the volume density of the buffer gas atoms.

The second effect originates from the interaction between the ion core of the Rydberg atom and the perturbers. When a Rydberg atom is immersed in a medium containing ground-state atoms or molecules, the atomic electric field induces a polarization in the perturbers, leading to a frequency shift. The frequency shift due to the polarization effect can be obtained from the impact approximation~\cite{alekseev1966spectroscopic,Omont1977}, and is given by (in units of rad/s)
\begin{equation}
\Delta\omega_{\text{p}} = -6.21
\left[\frac{\alpha e^{2}}{\hbar (4\pi\epsilon_0)^2}\right]^{2/3} v^{1/3} N,
\label{eq:delta_pol}
\end{equation}
where $\alpha$ represents the polarizability of the perturber in Cm$^2$/V, and $v$ is the mean relative velocity between the Rydberg atoms and the perturbers in m/s.
The total energy shift experienced by the Rydberg atom is the sum of both effects,
$\Delta \omega_{\text{total}} = \Delta\omega_{\text{sc}} + \Delta\omega_{\text{p}}.$

In addition to frequency shifts, polarization and electron scattering can also lead to level decays, denoted $\gamma_{\text{p}}$ and $\gamma_{\text{sc}}$, respectively. It was found that the level decay mainly comes from the polarization of the perturbing atoms~\cite{Omont1977}, {\sl{i.e.},} $\gamma_{\text{p}}>>\gamma_{\text{sc}}$, and that the decay rate 
\begin{equation}
\gamma_{\text{p}} = 2 \times 3.59
\left[\frac{\alpha e^{2}}{\hbar (4\pi\epsilon_0)^2}\right]^{2/3} v^{1/3} N \quad.
\label{eq:gamma_pol}
\end{equation}
The value of $\gamma_{\text{p}}$ is equivalent with a full-width at half-maximum (FWHM) line broadening in units of rad/s.

Through two-photon, Doppler-free spectroscopy, the broadening and shifts of Rb $nS$ and $nD$ Rydberg levels in the presence of inert perturbers has been experimentally observed in~\cite{Brillet1980} for He, Ar, Ne, Kr, and Xe, in~\cite{weber1982impact} for He, Ar, and Xe, in~\cite{bruce1982collision} for He and Ar, and in~\cite{Thompson1987} for Ne, Kr, and H$_2$. For these experiments, effects of buffer-gas pressure broadening by the intermediate $5P_{1/2}$ and $5P_{3/2}$ levels could be ignored because they were very far off-resonance. This is, however, not the case in our work. Pressure broadening of the Rb $D_2$ line due to binary interactions with noble gases has been observed in~\cite{Ottinger1975} for pressures of up to 1.1~kTorr. Recently, ultra-high pressures of He and Ar on the order of $10^{5}~$Torr interacting with a Rb vapor were spectroscopically studied in~\cite{ockenfels2022spectroscopy}. In the present work, the Ne pressure is 5~Torr, which leads to a $D_2$ line broadening on the order of $\gamma_{D_2} \sim 2\pi \times 50~$MHz~\cite{Ottinger1975}. This is consistent with our Doppler-free saturated absorption spectra, which only show marginal remnants of the buffer-gas-free $5P_{3/2}$ hyperfine features. Since we observe Rydberg-EIT linewidths that are considerably larger, at the level of precision of our current study we neglect the effect of $\gamma_{D_2}$ on the Rydberg-EIT linewidth. As such, based on 
$\gamma_{p} >> \gamma_{sc}$ and $\gamma_p > \gamma_{D_2}$, we compare our measured Rydberg-EIT linewidths only with estimates for $\gamma_{p}$.



\section{Experimental Setup}
\label{sec:exp}
\begin{figure*}[bthp]
\centering
\includegraphics[width=0.9\textwidth]{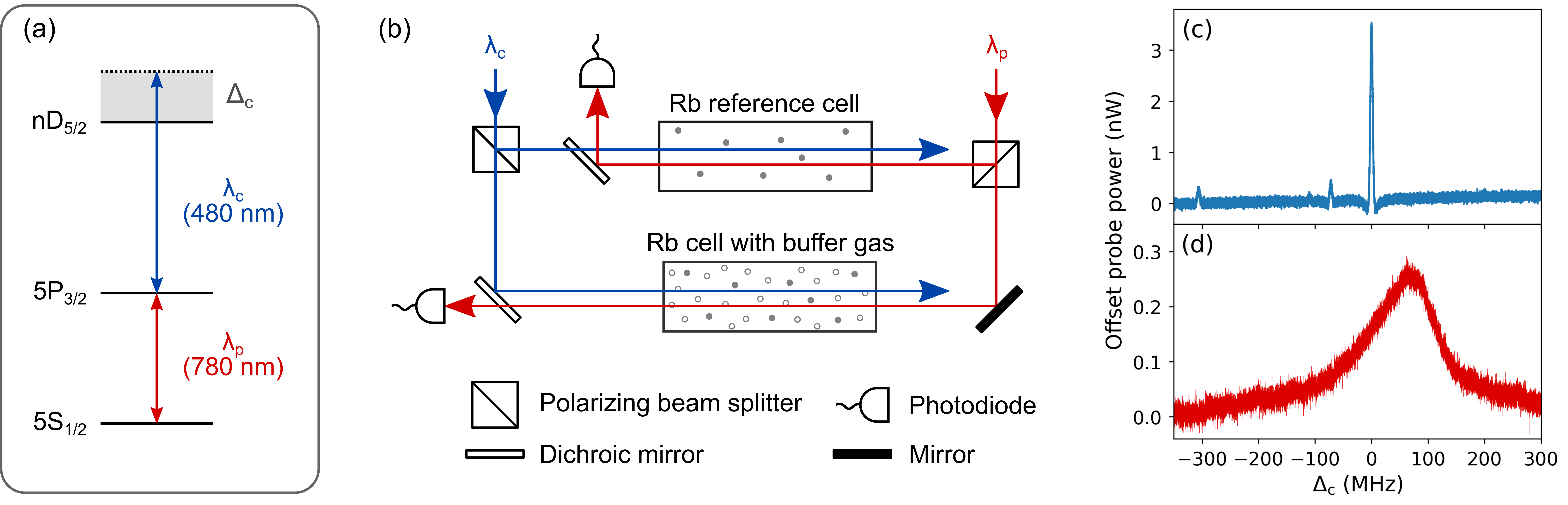}
\caption{(a) The energy-level diagram of rubidium atoms used in this work. The probe laser ($\lambda_{p}$) is on-resonance with the 5$S_{1/2}$~$\leftrightarrow$~5$P_{3/2}$ transition, while the coupling laser ($\lambda_{c}$) is detuned from the 5$P_{3/2}$~$\leftrightarrow$~n$D_{5/2}$ transition by $\Delta_{c}$. (b) An illustration of a Rydberg EIT setup. The EIT signals are detected simultaneously from both the Rb reference cell and the Rb cell with buffer gas. The 34D EIT signals from the Rb reference cell and the Rb cell with buffer gas are shown in (c) and (d), where the probe powers are 170~nW and 130~nW, respectively, and the coupler powers are approximately 2~mW and 35~mW.
}
\label{fig:expsetup}
\end{figure*}

We perform the EIT experiment using 
$^{85}$Rb atoms with an energy-level diagram shown in Fig.~\ref{fig:expsetup}~(a).
The probe laser ($\lambda_{\text{p}}$) with a wavelength of $\lambda_p=780$~nm is close to resonance with the $F=3$ to $F'=4$ hyperfine component of the 5$S_{1/2}$~$\leftrightarrow$~5$P_{3/2}$ transition. The EIT signal is measured as a function of the coupling laser (wavelength $\lambda_{\text{c}} \approx 480$~nm), 
which is scanned over the 5$P_{3/2}$~$\leftrightarrow$~$nD_{5/2}$ transition. The coupler detuning is denoted by $\Delta_{c}$.

We simultaneously extract EIT signals from two EIT beam lines, as shown in Fig.~\ref{fig:expsetup}~(b). The upper beam line, which serves to produce a reference spectrum, utilizes a buffer-gas-free Rb vapor cell. The reference EIT signal allows us to calibrate the frequency axis for $\Delta_c$, as well as to mark the coupler-laser frequency of the shift-free EIT line. In the lower (signal) beam line, the EIT signal is acquired using a cell that contains Rb vapor and a Ne buffer gas of nominally 5~Torr pressure. The probe and coupler lasers 
are split between the reference and signal beam lines using polarizing beam-splitter cubes (PBS), and are then counter-propagated through the respective cells. In each cell, both beams have parallel linear polarizations. After passage through the cells, probe and coupler beams are separated using dichroic optics. The  reference and signal probe beams are simultaneously detected using a pair of identical silicon photo-diodes and low-noise transimpedance amplifiers (TIAs), and the respective data traces are recorded.  We perform 100 scans per data set, and  present averages over the 100 scans.

The Rb vapor cell in the reference (upper) beam line in Fig.~\ref{fig:expsetup}~(c) is held at room temperature (291~K).
The probe and coupling beams in the reference line are approximately Gaussian  and have $1/e^2$ drop-off radii of the intensity distribution of $w_0=300~\mu$m and 500~$\mu$m,  
respectively. In the signal (lower) beam line, the cell that contains the buffer gas 
is heated to 303~K, which results in about 70\% peak absorption on the 5$S_{1/2}$, F$=3$ $\leftrightarrow $ 5$P_{3/2}$, F$'$ transition. The probe and coupling beams in the signal line have Gaussian beam parameters of $w_0 \approx 150~\mu$m. 

The reference EIT signal is shown in Fig.~\ref{fig:expsetup}~(c). The strongest EIT peak is from the 5$S_{1/2}$, $F=3$  $\leftrightarrow$ $5P_{3/2}$, $F'=4$ $\leftrightarrow$ 34$D_{5/2}$ cascade, which has the largest electric dipole moment and is the least diminished by optical pumping into the uncoupled 5$S_{1/2}$, $F=2$ level.   
The two small peaks that bracket the -100~MHz mark are from the intermediate hyperfine states 5$P_{3/2}$, $F'=3$ and
$F'=2$. These are small in size mostly due to optical pumping during the atom-field interaction time (which is a few $\mu$s in the buffer-gas-free cell). All observed frequency splittings between the $5P_{3/2}$ hyperfine peaks carry a Doppler scaling factor of $(\lambda_p/\lambda_c -1 ) = 0.63$. 
The leftmost peak in Fig.~\ref{fig:expsetup}~(c) is attributed to the 5$S_{1/2}$, $F=3$  $\leftrightarrow$ $5P_{3/2}$, $F'=4$ $\leftrightarrow$ 34$D_{3/2}$ cascade. Noting that the Doppler scaling factor for Rydberg lines is unity, the splitting between the largest 34$D_{5/2}$ peak and the 34$D_{3/2}$ peak in Fig.~\ref{fig:expsetup}~(c) equals the 34$D_{J}$ fine-structure splitting, which is 306.057~MHz. This splitting is used for calibration of the frequency axis of the reference and signal spectra, which are simultaneously acquired.

A typical Rydberg-EIT signal at a low probe intensity, obtained from the cell with buffer gas, is shown in Fig.~\ref{fig:expsetup}~(d). The EIT signal from that cell has an asymmetric shape, and in the case shown the peak is shifted positively from the reference EIT line 
by 68 $\pm$ 1~MHz.
In the following we study the dependence of shift and line width on principal quantum number $n$. We will then 
explore effects observed at high probe intensity.


\section{Frequency shifts and linewidths}
\label{sec:shifts}

\begin{figure}[t]
\centering
\includegraphics[width=0.84\linewidth]{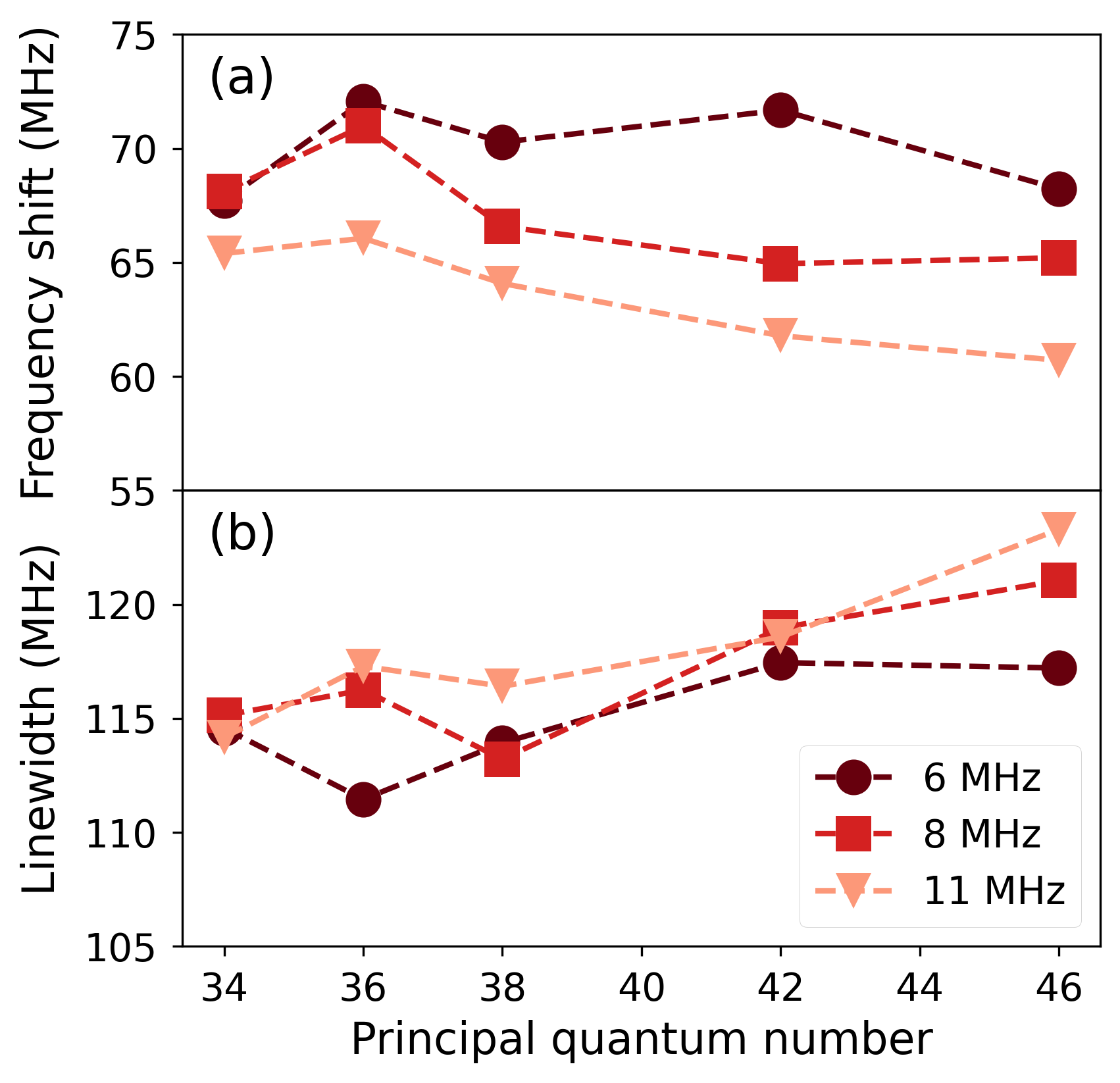}
\caption{(a) Frequency shift of the EIT signal from the Rb cell with buffer gas relative to the $nD_{5/2}$ peak for $F'=4$ in the reference EIT spectrum versus $n$, at several probe powers with estimated Rabi frequencies at the beam center shown in the legend. The plot data are extracted from local parabolic fits to the peaks in spectra. Fit error bars are smaller than the marker size. (b) Corresponding FWHM of the EIT peaks. The FWHM values are obtained from the roots of the first derivatives of smoothed EIT curves [see Fig.~\ref{fig:expsetup} (d) and Fig.~\ref{fig:vary_probe}]. Statistical uncertainties are smaller than the marker size.
}
\label{fig:peakfreq}
\end{figure}

To study the effect of the buffer gas on the EIT signals for a range of different Rydberg states, we take Rydberg-EIT data over an $n$-range of 34 to 46. We extract the relative frequency of the EIT peak in the signal beam, which equals the frequency shift of the EIT caused by the buffer gas. The frequency-shift
results are shown in Fig.~\ref{fig:peakfreq}~(a) for probe powers of 0.06, 0.13, and 0.22~$\mu$W, corresponding to the probe Rabi frequencies listed. The observed frequency shifts are in the range of 60-73~MHz and averages to about 67~MHz, with the low-power data clustering around 70~MHz. This result agrees well with the observations in~\cite{Thompson1987} where the shift rate was measured to be 12$\pm$1 MHz/Torr for high $n$ (which would lead to a shift of 60$\pm$5~MHz for 5~Torr of Ne buffer gas). Further, the shift decreases by up to about 10~MHz when the probe power increases, and it appears to overall decrease by a few MHz when $n$ increases. The weak $n$-dependence agrees with a semi-classical calculation in \cite{Henry2002}, which shows that the frequency shift of the Rydberg $nD$ state slightly decreases as $n$ increases. 

For a quantitative comparison with theory, we first calculate the electron-scattering shifts using Eq.~\ref{eq:delta_sc}. For the low-energy s-wave scattering length, $a_{s}$, several of the previously computed values include, from low to high, $a_s = 0.2 a_0$
~\cite{Hoffmann1969}, $0.227 a_0$~\cite{Fedus2014}, and $0.24 a_0$~\cite{Omont1977}. A table listing both theoretical and experimental values for $a_{s}$ is provided in \cite{Cheng2014}. Over the range $0.2 a_0 < a_{s} < 0.24 a_0$, 
$\Delta\omega_{\text{sc}}/(2 \pi)$ varies from 195~MHz to 234~MHz. 
The shift from the Ne polarization due to the Rydberg atom, obtained from Eq.~\ref{eq:delta_pol} and using $\alpha = 2.66$ in atomic units~\cite{Omont1977}, 
is $\Delta\omega_{\text{p}}/(2 \pi) = -122$~MHz. The net shift from these two effects, $\Delta \omega_{\text{total}}/(2 \pi)$, then ranges between 73~MHz and 112~MHz. It is seen that the shift from the calculation has the same sign in theory and experiment, confirming that the low-energy s-wave scattering length $a_s$ is positive (which is not the case for some other buffer gases) and dominant. Furthermore, depending on what exact value for $a_s$ is adopted, the calculated net shift is about 10\% to 60\% larger than the average experimental value of $\approx 67$~MHz from the previous paragraph. Hence, we claim good qualitative agreement. 

To discuss these findings, we first note that the overall spread of data previously reported for $a_s$ has the largest effect on the net frequency shift, $\Delta \omega_{\text{total}}$, with our result being more consistent with the lower end of previously reported $a_s$-values ($a_s \approx 0.2 a_0$). In fact, the ambiguity of estimates for 
$a_s$ alone may suffice to explain our observed deviation (if any) between measured line shifts and corresponding theoretical estimates. As to additional effects that might matter, we note that the Rydberg-EIT line may be slightly pulled to lower frequencies by the weak 
EIT peaks that are visible in Fig.~\ref{fig:expsetup}~(c) but that are hidden in Fig.~\ref{fig:expsetup}~(d). Furthermore, the Ne buffer gas pressure of 5~Torr has an uncertainty of 5\% according to manufacturer information. Along the same line, one may speculate that differences in cell temperature during cell fabrication and the cell's eventual operating temperature could in principle cause a mild buffer-gas density drop. 

Further progress on comparing experimental buffer-gas-induced Rydberg-EIT shifts and theoretical estimates would require a refinement of theoretical models for $a_s$ as well as a full model for Rydberg-EIT that covers the effects of the buffer gas on all involved atomic levels. In experimental work, one may consider a determination of the absolute Ne density with an independent, quantitative method. These research directions are, however, outside the scope of our present work.


We next extract the FWHM of the EIT signal from the cell with buffer gas for several probe powers and $n$-values. As shown in Fig.~\ref{fig:peakfreq}~(b), the observed FWHM of the EIT signal from the cell with the buffer gas is between 110 and 125~MHz, which greatly exceeds the width of the reference EIT lines. The FWHM of the buffer-gas EIT slightly increases with an increase in $n$, but it is not significantly dependent on the probe power. This agrees well with the calculation in \cite{Henry2002}, which shows that, for Rydberg $nD$ states, the FWHM of the Rydberg line should slightly increase with $n$. 
A calculation of the broadening $\gamma_p$ from the polarization effect (Eq.~\ref{eq:gamma_pol}) yields a FWHM of 144~MHz, which is $\approx 20\%$ larger than the experimentally observed width. We note that the effects that we neglect here could contribute somewhat to the experimentally observed line broadening, including broadening from the weak EIT peaks that are visible in Fig.~\ref{fig:expsetup}~(c) but that are hidden in Fig.~\ref{fig:expsetup}~(d), as well as the line broadening $\gamma_{D_2}$
[which is small compared to $\gamma_p$ but still substantially $>0$~(see Sec.~\ref{sec:theory})].
The results on the EIT linewidth may also indicate that the exact density of Ne atoms in the cell could be slightly lower than the density one would have at 5~Torr and at room temperature.


\section{Effects of probe power}

\begin{figure}[t]
\centering
\includegraphics[width=\linewidth]{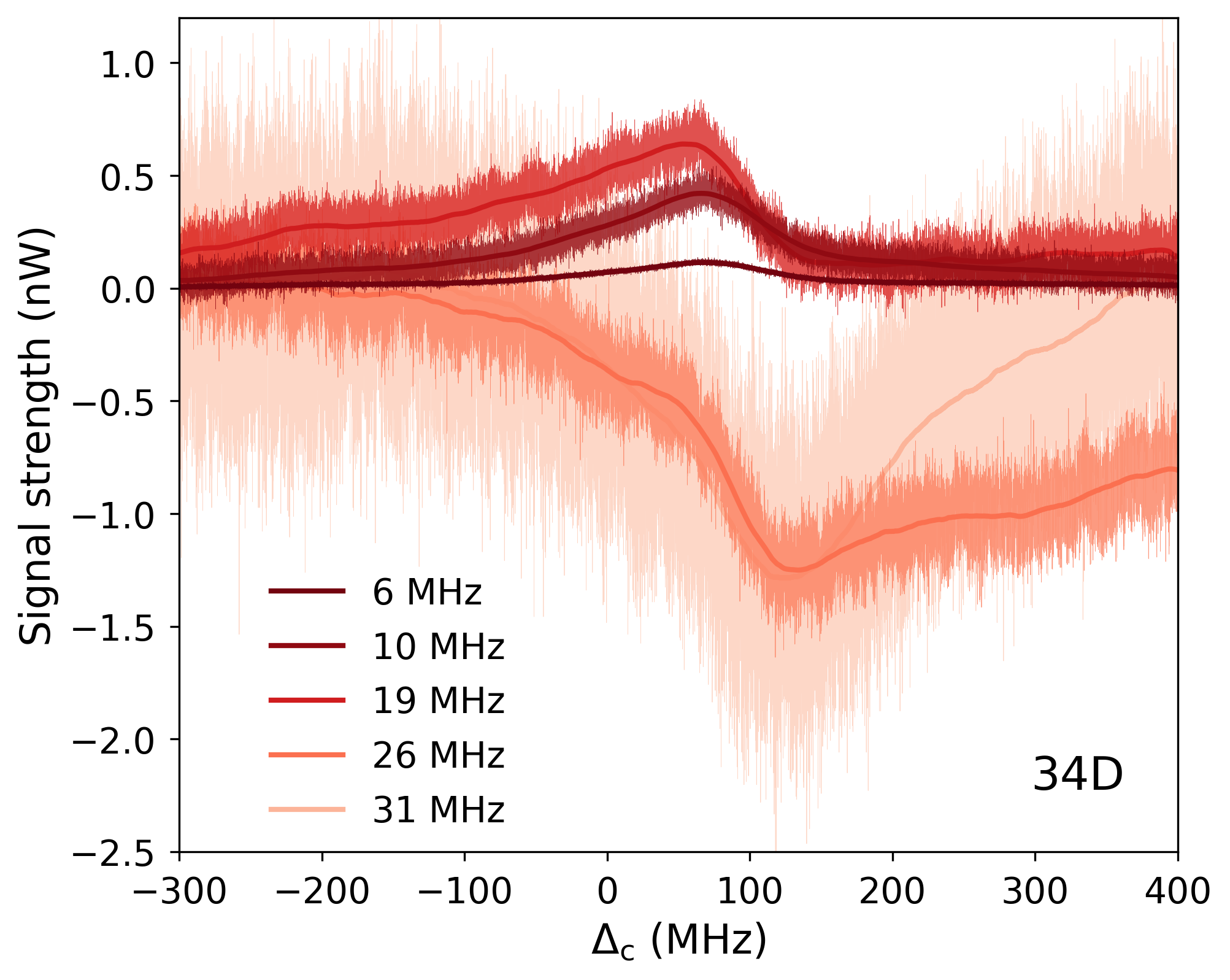}
\caption{EIT signals of  Rb~34$D_{5/2}$ Rydberg atoms in the cell with buffer gas for probe laser powers ranging  from  0.06 to 1.79~$\mu$W. The colors of the plots correspond to probe-laser Rabi frequency, $\Omega_p/(2 \pi)$, indicated in the legend. To allow for a comparison of the signals on a fixed scale for the transmitted probe power, the plots are vertically shifted so that they level out at zero at large detunings. Also, the TIA gain is decreased with increasing probe power to avoid saturation. The plot shows data at the full TIA bandwidth as well as smoothed curves that allow for an easier comparison of the atomic response across the full probe-power range.}
\label{fig:vary_probe}
\end{figure}

Finally, we study the effects of probe power on the Rydberg-EIT signal in the 
signal beam line, which harbors the rubidium cell with the buffer gas, by varying the probe power from $P_p = 0.06~\mu$W to 1.79~$\mu$W. The central probe-laser electric field is $E_p = \sqrt{2 I_p / (c \epsilon_0)}$, with central intensity $I_p = 2 P_p/(\pi w_0^2)$ and $w_0 \approx 150~\mu$m for the signal beam line. The probe Rabi frequency $\Omega_p/(2 \pi) = \mu_{12} E_p / h$, where $\mu_{12} = 1.892 \, e \, a_0$~\cite{Steck} is the probe-transition dipole moment.
The EIT signals from the 34$D_{5/2}$ state are shown in Fig.~\ref{fig:vary_probe} for five values of the probe Rabi frequency. The noise increase at high probe power is attributed to an increase in shot noise as well as an increase in TIA bandwidth at lower gain. Over the gain values used, the bandwidth of the TIA (model SRS SR570) increases from 200~Hz to 2~kHz with increasing probe power. Hence, the signals at higher probe powers have considerably larger noise on the utilized absolute-power scale. In Fig.~\ref{fig:vary_probe}, we include smoothed curves that allow for an easier comparison of the signal behavior over the entire probe-power range.

It can be seen in Fig.~\ref{fig:vary_probe} that at low probe powers (Rabi frequency $\lesssim 10$~MHz, as used in Sec.~\ref{sec:shifts}) the EIT 
line shape is invariant and the EIT 
signal strength is in proportion with probe power. In the limit of vanishing probe power
such a linear behavior is expected. 
Also, the shape of the EIT signal is asymmetric, with a longer tail on the negative side. We speculate that this behavior may come in part from the blending of intermediate-state hyperfine structure, which adds to the $nD$-line broadening \cite{Ottinger1975}. At Rabi frequencies above $\sim$10~MHz, the EIT peak position begins to shift negative, while shape and width of the peak still largely remain the same.  At probe Rabi frequencies exceeding $\sim$20~MHz, the signals invert in shape and turn into electromagnetically-induced absorption (EIA), with the center of the EIA dip located $\sim 50$~MHz above the low-power EIT peak. The transition from EIT to EIA at high power may come from factors that involve optical pumping and velocity-changing collisions, the study of which could be the subject of future work.


\section{Conclusion}

We have observed Rydberg-EIT in a vapor cell containing 5~Torr of Ne buffer gas. Results obtained at low probe power have revealed frequency shifts of the EIT signals by about 70~MHz, as well as an increased FWHM EIT linewidth of about 120~MHz. These observations are largely unaffected by variations in the principal quantum number of the Rydberg states and in the probe power, as long as the probe Rabi frequency remains below about 10~MHz. The frequency shift is attributed to low-energy s-wave scattering between the Rydberg electron and the Ne atoms and polarization of there Ne atoms by the atomic electric field. The width of the signal is dominated by polarization of the Ne atoms. At high probe power, we observe a transition from EIT to EIA; this phenomenon awaits a future explanation.  

Utilizing the Stark effect of Rydberg levels, potential applications of our research include non-invasive and spatially-resolved measurement of electric fields in low-pressure discharge plasmas in neon. Further, the electric fields of highly-charged dust particles in low-density plasma can potentially be mapped via Rydberg-EIT. For pressures below 
about 5~Torr, we also see applications in using Rydberg-EIT as a real-time, in-situ and non-invasive readout for buffer-gas density at a location of interest, which can have advantages over reading the buffer-gas pressure with a remote pressure gauge.

\section*{Acknowledgments}
\label{sec:acknowledgments}
We acknowledge fruitful discussions with Prof. Eric Paradis (Eastern Michigan University), Dr. David A. Anderson (Rydberg Technologies Inc.), and Bineet Dash (University of Michigan).  This project was supported by the U.S. Department of Energy, Office of Science, Office of Fusion Energy Sciences under award number DE-SC0023090.
N.T. acknowledges funding from the NSRF via the Program Management Unit for Human Resources and Institutional Development, Research and Innovation (grant number B05F650024), and from the Office of the Permanent Secretary, Ministry of Higher Education, Science, Research and Innovation (Grant No. RGNS.64-067). R.C. acknowledges support from a Rackham Predoctoral Fellowship of the University of Michigan.

\bibliography{references.bib}

\begin{thebibliography}{43}%
\makeatletter
\providecommand \@ifxundefined [1]{%
 \@ifx{#1\undefined}
}%
\providecommand \@ifnum [1]{%
 \ifnum #1\expandafter \@firstoftwo
 \else \expandafter \@secondoftwo
 \fi
}%
\providecommand \@ifx [1]{%
 \ifx #1\expandafter \@firstoftwo
 \else \expandafter \@secondoftwo
 \fi
}%
\providecommand \natexlab [1]{#1}%
\providecommand \enquote  [1]{``#1''}%
\providecommand \bibnamefont  [1]{#1}%
\providecommand \bibfnamefont [1]{#1}%
\providecommand \citenamefont [1]{#1}%
\providecommand \href@noop [0]{\@secondoftwo}%
\providecommand \href [0]{\begingroup \@sanitize@url \@href}%
\providecommand \@href[1]{\@@startlink{#1}\@@href}%
\providecommand \@@href[1]{\endgroup#1\@@endlink}%
\providecommand \@sanitize@url [0]{\catcode `\\12\catcode `\$12\catcode
  `\&12\catcode `\#12\catcode `\^12\catcode `\_12\catcode `\%12\relax}%
\providecommand \@@startlink[1]{}%
\providecommand \@@endlink[0]{}%
\providecommand \url  [0]{\begingroup\@sanitize@url \@url }%
\providecommand \@url [1]{\endgroup\@href {#1}{\urlprefix }}%
\providecommand \urlprefix  [0]{URL }%
\providecommand \Eprint [0]{\href }%
\providecommand \doibase [0]{http://dx.doi.org/}%
\providecommand \selectlanguage [0]{\@gobble}%
\providecommand \bibinfo  [0]{\@secondoftwo}%
\providecommand \bibfield  [0]{\@secondoftwo}%
\providecommand \translation [1]{[#1]}%
\providecommand \BibitemOpen [0]{}%
\providecommand \bibitemStop [0]{}%
\providecommand \bibitemNoStop [0]{.\EOS\space}%
\providecommand \EOS [0]{\spacefactor3000\relax}%
\providecommand \BibitemShut  [1]{\csname bibitem#1\endcsname}%
\let\auto@bib@innerbib\@empty
\bibitem [{\citenamefont {Sedlacek}\ \emph {et~al.}(2012)\citenamefont
  {Sedlacek}, \citenamefont {Schwettmann}, \citenamefont {K\"ubler},
  \citenamefont {L\"ow}, \citenamefont {Pfau},\ and\ \citenamefont
  {Shaffer}}]{sedlacek12}%
  \BibitemOpen
  \bibfield  {author} {\bibinfo {author} {\bibfnamefont {J.~A.}\ \bibnamefont
  {Sedlacek}}, \bibinfo {author} {\bibfnamefont {A.}~\bibnamefont
  {Schwettmann}}, \bibinfo {author} {\bibfnamefont {H.}~\bibnamefont
  {K\"ubler}}, \bibinfo {author} {\bibfnamefont {R.}~\bibnamefont {L\"ow}},
  \bibinfo {author} {\bibfnamefont {T.}~\bibnamefont {Pfau}}, \ and\ \bibinfo
  {author} {\bibfnamefont {J.~P.}\ \bibnamefont {Shaffer}},\ }\bibfield
  {title} {\enquote {\bibinfo {title} {Microwave electrometry with rydberg
  atoms in a vapour cell using bright atomic resonances},}\ }\href {\doibase
  10.1038/nphys2423} {\bibfield  {journal} {\bibinfo  {journal} {Nature
  Physics}\ }\textbf {\bibinfo {volume} {8}},\ \bibinfo {pages} {819} (\bibinfo
  {year} {2012})}\BibitemShut {NoStop}%
\bibitem [{\citenamefont {Sedlacek}\ \emph {et~al.}(2013)\citenamefont
  {Sedlacek}, \citenamefont {Schwettmann}, \citenamefont {K\"ubler},\ and\
  \citenamefont {Shaffer}}]{sedlacek13}%
  \BibitemOpen
  \bibfield  {author} {\bibinfo {author} {\bibfnamefont {J.~A.}\ \bibnamefont
  {Sedlacek}}, \bibinfo {author} {\bibfnamefont {A.}~\bibnamefont
  {Schwettmann}}, \bibinfo {author} {\bibfnamefont {H.}~\bibnamefont
  {K\"ubler}}, \ and\ \bibinfo {author} {\bibfnamefont {J.~P.}\ \bibnamefont
  {Shaffer}},\ }\bibfield  {title} {\enquote {\bibinfo {title} {Atom-based
  vector microwave electrometry using rubidium rydberg atoms in a vapor
  cell},}\ }\href {\doibase 10.1103/PhysRevLett.111.063001} {\bibfield
  {journal} {\bibinfo  {journal} {Phys. Rev. Lett.}\ }\textbf {\bibinfo
  {volume} {111}},\ \bibinfo {pages} {063001} (\bibinfo {year}
  {2013})}\BibitemShut {NoStop}%
\bibitem [{\citenamefont {{Holloway}}\ \emph {et~al.}(2014)\citenamefont
  {{Holloway}}, \citenamefont {{Gordon}}, \citenamefont {{Jefferts}},
  \citenamefont {{Schwarzkopf}}, \citenamefont {{Anderson}}, \citenamefont
  {{Miller}}, \citenamefont {{Thaicharoen}},\ and\ \citenamefont
  {{Raithel}}}]{holloway14}%
  \BibitemOpen
  \bibfield  {author} {\bibinfo {author} {\bibfnamefont {C.~L.}\ \bibnamefont
  {{Holloway}}}, \bibinfo {author} {\bibfnamefont {J.~A.}\ \bibnamefont
  {{Gordon}}}, \bibinfo {author} {\bibfnamefont {S.}~\bibnamefont
  {{Jefferts}}}, \bibinfo {author} {\bibfnamefont {A.}~\bibnamefont
  {{Schwarzkopf}}}, \bibinfo {author} {\bibfnamefont {D.~A.}\ \bibnamefont
  {{Anderson}}}, \bibinfo {author} {\bibfnamefont {S.~A.}\ \bibnamefont
  {{Miller}}}, \bibinfo {author} {\bibfnamefont {N.}~\bibnamefont
  {{Thaicharoen}}}, \ and\ \bibinfo {author} {\bibfnamefont {G.}~\bibnamefont
  {{Raithel}}},\ }\bibfield  {title} {\enquote {\bibinfo {title} {Broadband
  rydberg atom-based electric-field probe for si-traceable, self-calibrated
  measurements},}\ }\href {\doibase 10.1109/TAP.2014.2360208} {\bibfield
  {journal} {\bibinfo  {journal} {IEEE Transactions on Antennas and
  Propagation}\ }\textbf {\bibinfo {volume} {62}},\ \bibinfo {pages}
  {6169--6182} (\bibinfo {year} {2014})}\BibitemShut {NoStop}%
\bibitem [{\citenamefont {Anderson}\ \emph {et~al.}(2020)\citenamefont
  {Anderson}, \citenamefont {Sapiro},\ and\ \citenamefont
  {Raithel}}]{Anderson2020}%
  \BibitemOpen
  \bibfield  {author} {\bibinfo {author} {\bibfnamefont {D.A.}\ \bibnamefont
  {Anderson}}, \bibinfo {author} {\bibfnamefont {R.E.}\ \bibnamefont {Sapiro}},
  \ and\ \bibinfo {author} {\bibfnamefont {G.}~\bibnamefont {Raithel}},\
  }\bibfield  {title} {\enquote {\bibinfo {title} {Rydberg atoms for
  radio-frequency communications and sensing: atomic receivers for pulsed rf
  field and phase detection},}\ }\href
  {http://dx.doi.org/10.1109/MAES.2019.2960922} {\bibfield  {journal} {\bibinfo
   {journal} {IEEE Aerosp. Electron. Syst. Mag.}\ }\textbf {\bibinfo {volume}
  {35}},\ \bibinfo {pages} {48 -- 56} (\bibinfo {year} {2020})}\BibitemShut
  {NoStop}%
\bibitem [{\citenamefont {Meyer}\ \emph {et~al.}(2021)\citenamefont {Meyer},
  \citenamefont {Kunz},\ and\ \citenamefont {Cox}}]{Meyer2021}%
  \BibitemOpen
  \bibfield  {author} {\bibinfo {author} {\bibfnamefont {David~H.}\
  \bibnamefont {Meyer}}, \bibinfo {author} {\bibfnamefont {Paul~D.}\
  \bibnamefont {Kunz}}, \ and\ \bibinfo {author} {\bibfnamefont {Kevin~C.}\
  \bibnamefont {Cox}},\ }\bibfield  {title} {\enquote {\bibinfo {title}
  {Waveguide-coupled rydberg spectrum analyzer from 0 to 20 ghz},}\ }\href
  {\doibase 10.1103/PhysRevApplied.15.014053} {\bibfield  {journal} {\bibinfo
  {journal} {Phys. Rev. Appl.}\ }\textbf {\bibinfo {volume} {15}},\ \bibinfo
  {pages} {014053} (\bibinfo {year} {2021})}\BibitemShut {NoStop}%
\bibitem [{\citenamefont {Anderson}\ \emph {et~al.}(2021)\citenamefont
  {Anderson}, \citenamefont {Sapiro},\ and\ \citenamefont
  {Raithel}}]{receiver2021}%
  \BibitemOpen
  \bibfield  {author} {\bibinfo {author} {\bibfnamefont {David~Alexander}\
  \bibnamefont {Anderson}}, \bibinfo {author} {\bibfnamefont
  {Rachel~Elizabeth}\ \bibnamefont {Sapiro}}, \ and\ \bibinfo {author}
  {\bibfnamefont {Georg}\ \bibnamefont {Raithel}},\ }\bibfield  {title}
  {\enquote {\bibinfo {title} {An atomic receiver for am and fm radio
  communication},}\ }\href {\doibase 10.1109/TAP.2020.2987112} {\bibfield
  {journal} {\bibinfo  {journal} {IEEE Trans. Antennas Propag.}\ }\textbf
  {\bibinfo {volume} {69}},\ \bibinfo {pages} {2455--2462} (\bibinfo {year}
  {2021})}\BibitemShut {NoStop}%
\bibitem [{\citenamefont {Feldbaum}\ \emph {et~al.}(2002)\citenamefont
  {Feldbaum}, \citenamefont {Morrow}, \citenamefont {Dutta},\ and\
  \citenamefont {Raithel}}]{Feldbaum2002}%
  \BibitemOpen
  \bibfield  {author} {\bibinfo {author} {\bibfnamefont {D.}~\bibnamefont
  {Feldbaum}}, \bibinfo {author} {\bibfnamefont {N.~V.}\ \bibnamefont
  {Morrow}}, \bibinfo {author} {\bibfnamefont {S.~K.}\ \bibnamefont {Dutta}}, \
  and\ \bibinfo {author} {\bibfnamefont {G.}~\bibnamefont {Raithel}},\
  }\bibfield  {title} {\enquote {\bibinfo {title} {Coulomb expansion of
  laser-excited ion plasmas},}\ }\href {\doibase 10.1103/PhysRevLett.89.173004}
  {\bibfield  {journal} {\bibinfo  {journal} {Phys. Rev. Lett.}\ }\textbf
  {\bibinfo {volume} {89}},\ \bibinfo {pages} {173004} (\bibinfo {year}
  {2002})}\BibitemShut {NoStop}%
\bibitem [{\citenamefont {Duspayev}\ and\ \citenamefont
  {Raithel}(2023)}]{Duspayev2023}%
  \BibitemOpen
  \bibfield  {author} {\bibinfo {author} {\bibfnamefont {Alisher}\ \bibnamefont
  {Duspayev}}\ and\ \bibinfo {author} {\bibfnamefont {Georg}\ \bibnamefont
  {Raithel}},\ }\bibfield  {title} {\enquote {\bibinfo {title} {Electric field
  analysis in a cold-ion source using stark spectroscopy of rydberg atoms},}\
  }\href {\doibase 10.1103/PhysRevApplied.19.044051} {\bibfield  {journal}
  {\bibinfo  {journal} {Phys. Rev. Appl.}\ }\textbf {\bibinfo {volume} {19}},\
  \bibinfo {pages} {044051} (\bibinfo {year} {2023})}\BibitemShut {NoStop}%
\bibitem [{\citenamefont {Anderson}\ \emph {et~al.}(2017)\citenamefont
  {Anderson}, \citenamefont {Raithel}, \citenamefont {Simons},\ and\
  \citenamefont {Holloway}}]{anderson17}%
  \BibitemOpen
  \bibfield  {author} {\bibinfo {author} {\bibfnamefont {David~A.}\
  \bibnamefont {Anderson}}, \bibinfo {author} {\bibfnamefont {Georg}\
  \bibnamefont {Raithel}}, \bibinfo {author} {\bibfnamefont {Matthew}\
  \bibnamefont {Simons}}, \ and\ \bibinfo {author} {\bibfnamefont
  {Christopher~L.}\ \bibnamefont {Holloway}},\ }\href@noop {} {\enquote
  {\bibinfo {title} {Quantum-optical spectroscopy for plasma electric field
  measurements and diagnostics},}\ } (\bibinfo {year} {2017}),\ \Eprint
  {http://arxiv.org/abs/1712.08717} {arXiv:1712.08717 [physics.atom-ph]}
  \BibitemShut {NoStop}%
\bibitem [{\citenamefont {Weller}\ \emph {et~al.}(2019)\citenamefont {Weller},
  \citenamefont {Shaffer}, \citenamefont {Pfau}, \citenamefont {L\"ow},\ and\
  \citenamefont {K\"ubler}}]{weller19}%
  \BibitemOpen
  \bibfield  {author} {\bibinfo {author} {\bibfnamefont {Daniel}\ \bibnamefont
  {Weller}}, \bibinfo {author} {\bibfnamefont {James~P.}\ \bibnamefont
  {Shaffer}}, \bibinfo {author} {\bibfnamefont {Tilman}\ \bibnamefont {Pfau}},
  \bibinfo {author} {\bibfnamefont {Robert}\ \bibnamefont {L\"ow}}, \ and\
  \bibinfo {author} {\bibfnamefont {Harald}\ \bibnamefont {K\"ubler}},\
  }\bibfield  {title} {\enquote {\bibinfo {title} {Interplay between thermal
  rydberg gases and plasmas},}\ }\href {\doibase 10.1103/PhysRevA.99.043418}
  {\bibfield  {journal} {\bibinfo  {journal} {Phys. Rev. A}\ }\textbf {\bibinfo
  {volume} {99}},\ \bibinfo {pages} {043418} (\bibinfo {year}
  {2019})}\BibitemShut {NoStop}%
\bibitem [{\citenamefont {Ma}\ \emph {et~al.}(2020)\citenamefont {Ma},
  \citenamefont {Paradis},\ and\ \citenamefont {Raithel}}]{ma2020dc}%
  \BibitemOpen
  \bibfield  {author} {\bibinfo {author} {\bibfnamefont {Lu}~\bibnamefont
  {Ma}}, \bibinfo {author} {\bibfnamefont {Eric}\ \bibnamefont {Paradis}}, \
  and\ \bibinfo {author} {\bibfnamefont {Georg}\ \bibnamefont {Raithel}},\
  }\bibfield  {title} {\enquote {\bibinfo {title} {Dc electric fields in
  electrode-free glass vapor cell by photoillumination},}\ }\href@noop {}
  {\bibfield  {journal} {\bibinfo  {journal} {Optics Express}\ }\textbf
  {\bibinfo {volume} {28}},\ \bibinfo {pages} {3676--3685} (\bibinfo {year}
  {2020})}\BibitemShut {NoStop}%
\bibitem [{\citenamefont {Bell}\ \emph {et~al.}(1961)\citenamefont {Bell},
  \citenamefont {Bloom},\ and\ \citenamefont {Lynch}}]{bell1961alkali}%
  \BibitemOpen
  \bibfield  {author} {\bibinfo {author} {\bibfnamefont {William~E}\
  \bibnamefont {Bell}}, \bibinfo {author} {\bibfnamefont {Arnold~L}\
  \bibnamefont {Bloom}}, \ and\ \bibinfo {author} {\bibfnamefont {James}\
  \bibnamefont {Lynch}},\ }\bibfield  {title} {\enquote {\bibinfo {title}
  {Alkali metal vapor spectral lamps},}\ }\href@noop {} {\bibfield  {journal}
  {\bibinfo  {journal} {Review of Scientific Instruments}\ }\textbf {\bibinfo
  {volume} {32}},\ \bibinfo {pages} {688--692} (\bibinfo {year}
  {1961})}\BibitemShut {NoStop}%
\bibitem [{\citenamefont {Brewer}(1961)}]{brewer1961high}%
  \BibitemOpen
  \bibfield  {author} {\bibinfo {author} {\bibfnamefont {Richard~G}\
  \bibnamefont {Brewer}},\ }\bibfield  {title} {\enquote {\bibinfo {title}
  {High intensity low noise rubidium light source},}\ }\href@noop {} {\bibfield
   {journal} {\bibinfo  {journal} {Review of Scientific Instruments}\ }\textbf
  {\bibinfo {volume} {32}},\ \bibinfo {pages} {1356--1358} (\bibinfo {year}
  {1961})}\BibitemShut {NoStop}%
\bibitem [{\citenamefont {Shukla}\ and\ \citenamefont
  {Eliasson}(2009)}]{Shukla2009}%
  \BibitemOpen
  \bibfield  {author} {\bibinfo {author} {\bibfnamefont {P.~K.}\ \bibnamefont
  {Shukla}}\ and\ \bibinfo {author} {\bibfnamefont {B.}~\bibnamefont
  {Eliasson}},\ }\bibfield  {title} {\enquote {\bibinfo {title} {Colloquium:
  Fundamentals of dust-plasma interactions},}\ }\href {\doibase
  10.1103/RevModPhys.81.25} {\bibfield  {journal} {\bibinfo  {journal} {Rev.
  Mod. Phys.}\ }\textbf {\bibinfo {volume} {81}},\ \bibinfo {pages} {25--44}
  (\bibinfo {year} {2009})}\BibitemShut {NoStop}%
\bibitem [{\citenamefont {Mitchell}\ \emph {et~al.}(2006)\citenamefont
  {Mitchell}, \citenamefont {Hor{\'a}nyi}, \citenamefont {Havnes},\ and\
  \citenamefont {Porco}}]{mitchell2006saturn}%
  \BibitemOpen
  \bibfield  {author} {\bibinfo {author} {\bibfnamefont {C~J}\ \bibnamefont
  {Mitchell}}, \bibinfo {author} {\bibfnamefont {M}~\bibnamefont
  {Hor{\'a}nyi}}, \bibinfo {author} {\bibfnamefont {O}~\bibnamefont {Havnes}},
  \ and\ \bibinfo {author} {\bibfnamefont {C~C}\ \bibnamefont {Porco}},\
  }\bibfield  {title} {\enquote {\bibinfo {title} {Saturn's spokes: Lost and
  found},}\ }\href@noop {} {\bibfield  {journal} {\bibinfo  {journal}
  {Science}\ }\textbf {\bibinfo {volume} {311}},\ \bibinfo {pages} {1587--1589}
  (\bibinfo {year} {2006})}\BibitemShut {NoStop}%
\bibitem [{\citenamefont {Renno}\ \emph {et~al.}(2003)\citenamefont {Renno},
  \citenamefont {Wong}, \citenamefont {Atreya}, \citenamefont {de~Pater},\ and\
  \citenamefont {Roos-Serote}}]{renno2003electrical}%
  \BibitemOpen
  \bibfield  {author} {\bibinfo {author} {\bibfnamefont {Nilton~O}\
  \bibnamefont {Renno}}, \bibinfo {author} {\bibfnamefont {Ah-San}\
  \bibnamefont {Wong}}, \bibinfo {author} {\bibfnamefont {Sushil~K}\
  \bibnamefont {Atreya}}, \bibinfo {author} {\bibfnamefont {Imke}\ \bibnamefont
  {de~Pater}}, \ and\ \bibinfo {author} {\bibfnamefont {Maarten}\ \bibnamefont
  {Roos-Serote}},\ }\bibfield  {title} {\enquote {\bibinfo {title} {Electrical
  discharges and broadband radio emission by martian dust devils and dust
  storms},}\ }\href@noop {} {\bibfield  {journal} {\bibinfo  {journal}
  {Geophysical Research Letters}\ }\textbf {\bibinfo {volume} {30}} (\bibinfo
  {year} {2003})}\BibitemShut {NoStop}%
\bibitem [{\citenamefont {Popel}\ \emph {et~al.}(2018)\citenamefont {Popel},
  \citenamefont {Zelenyi}, \citenamefont {Golub'},\ and\ \citenamefont
  {Dubinskii}}]{Popel2018}%
  \BibitemOpen
  \bibfield  {author} {\bibinfo {author} {\bibfnamefont {S.I.}\ \bibnamefont
  {Popel}}, \bibinfo {author} {\bibfnamefont {L.M.}\ \bibnamefont {Zelenyi}},
  \bibinfo {author} {\bibfnamefont {A.P.}\ \bibnamefont {Golub'}}, \ and\
  \bibinfo {author} {\bibfnamefont {A.Yu.}\ \bibnamefont {Dubinskii}},\
  }\bibfield  {title} {\enquote {\bibinfo {title} {Lunar dust and dusty
  plasmas: Recent developments, advances, and unsolved problems},}\ }\href
  {\doibase https://doi.org/10.1016/j.pss.2018.02.010} {\bibfield  {journal}
  {\bibinfo  {journal} {Planetary and Space Science}\ }\textbf {\bibinfo
  {volume} {156}},\ \bibinfo {pages} {71--84} (\bibinfo {year}
  {2018})}\BibitemShut {NoStop}%
\bibitem [{\citenamefont {Ratynskaia}\ \emph {et~al.}(2022)\citenamefont
  {Ratynskaia}, \citenamefont {Bortolon},\ and\ \citenamefont
  {Krasheninnikov}}]{Ratynskaia2022}%
  \BibitemOpen
  \bibfield  {author} {\bibinfo {author} {\bibfnamefont {S.}~\bibnamefont
  {Ratynskaia}}, \bibinfo {author} {\bibfnamefont {A.}~\bibnamefont
  {Bortolon}}, \ and\ \bibinfo {author} {\bibfnamefont {S.~I.}\ \bibnamefont
  {Krasheninnikov}},\ }\bibfield  {title} {\enquote {\bibinfo {title} {Dust and
  powder in fusion plasmas: recent developments in theory, modeling, and
  experiments},}\ }\href {\doibase 10.1007/s41614-022-00081-5} {\bibfield
  {journal} {\bibinfo  {journal} {Rev. Mod. Plasma Phys.}\ }\textbf {\bibinfo
  {volume} {6}},\ \bibinfo {pages} {20} (\bibinfo {year} {2022})}\BibitemShut
  {NoStop}%
\bibitem [{\citenamefont {Boufendi}\ \emph {et~al.}(2011)\citenamefont
  {Boufendi}, \citenamefont {Jouanny}, \citenamefont {Kovacevic}, \citenamefont
  {Berndt},\ and\ \citenamefont {Mikikian}}]{Boufendi2011}%
  \BibitemOpen
  \bibfield  {author} {\bibinfo {author} {\bibfnamefont {Laïfa}\ \bibnamefont
  {Boufendi}}, \bibinfo {author} {\bibfnamefont {M}~\bibnamefont {Jouanny}},
  \bibinfo {author} {\bibfnamefont {Eva}\ \bibnamefont {Kovacevic}}, \bibinfo
  {author} {\bibfnamefont {Johannes}\ \bibnamefont {Berndt}}, \ and\ \bibinfo
  {author} {\bibfnamefont {Maxime}\ \bibnamefont {Mikikian}},\ }\bibfield
  {title} {\enquote {\bibinfo {title} {Dusty plasma for nanotechnology},}\
  }\href {\doibase 10.1088/0022-3727/44/17/174035} {\bibfield  {journal}
  {\bibinfo  {journal} {Journal of Physics D: Applied Physics}\ }\textbf
  {\bibinfo {volume} {44}},\ \bibinfo {pages} {174035} (\bibinfo {year}
  {2011})}\BibitemShut {NoStop}%
\bibitem [{\citenamefont {Merlino}(2009)}]{merlino2009dust}%
  \BibitemOpen
  \bibfield  {author} {\bibinfo {author} {\bibfnamefont {Robert~L}\
  \bibnamefont {Merlino}},\ }\bibfield  {title} {\enquote {\bibinfo {title}
  {Dust-acoustic waves driven by an ion-dust streaming instability in
  laboratory discharge dusty plasma experiments},}\ }\href@noop {} {\bibfield
  {journal} {\bibinfo  {journal} {Physics of Plasmas}\ }\textbf {\bibinfo
  {volume} {16}},\ \bibinfo {pages} {124501} (\bibinfo {year}
  {2009})}\BibitemShut {NoStop}%
\bibitem [{\citenamefont {Menati}\ \emph {et~al.}(2019)\citenamefont {Menati},
  \citenamefont {Thomas},\ and\ \citenamefont
  {Kushner}}]{menati2019filamentation}%
  \BibitemOpen
  \bibfield  {author} {\bibinfo {author} {\bibfnamefont {Mohamad}\ \bibnamefont
  {Menati}}, \bibinfo {author} {\bibfnamefont {Edward}\ \bibnamefont {Thomas}},
  \ and\ \bibinfo {author} {\bibfnamefont {Mark~J}\ \bibnamefont {Kushner}},\
  }\bibfield  {title} {\enquote {\bibinfo {title} {Filamentation of
  capacitively coupled plasmas in large magnetic fields},}\ }\href@noop {}
  {\bibfield  {journal} {\bibinfo  {journal} {Physics of Plasmas}\ }\textbf
  {\bibinfo {volume} {26}},\ \bibinfo {pages} {063515} (\bibinfo {year}
  {2019})}\BibitemShut {NoStop}%
\bibitem [{\citenamefont {Thoma}\ \emph {et~al.}(2023)\citenamefont {Thoma},
  \citenamefont {Thomas}, \citenamefont {Knapek}, \citenamefont {Melzer},\ and\
  \citenamefont {Konopka}}]{thoma2023complex}%
  \BibitemOpen
  \bibfield  {author} {\bibinfo {author} {\bibfnamefont {Markus~H}\
  \bibnamefont {Thoma}}, \bibinfo {author} {\bibfnamefont {Hubertus~M}\
  \bibnamefont {Thomas}}, \bibinfo {author} {\bibfnamefont {Christina~A}\
  \bibnamefont {Knapek}}, \bibinfo {author} {\bibfnamefont {Andre}\
  \bibnamefont {Melzer}}, \ and\ \bibinfo {author} {\bibfnamefont {Uwe}\
  \bibnamefont {Konopka}},\ }\bibfield  {title} {\enquote {\bibinfo {title}
  {Complex plasma research under microgravity conditions},}\ }\href@noop {}
  {\bibfield  {journal} {\bibinfo  {journal} {npj Microgravity}\ }\textbf
  {\bibinfo {volume} {9}},\ \bibinfo {pages} {13} (\bibinfo {year}
  {2023})}\BibitemShut {NoStop}%
\bibitem [{\citenamefont {Sheridan}\ \emph {et~al.}(2016)\citenamefont
  {Sheridan}, \citenamefont {Weiner},\ and\ \citenamefont
  {Steinberger}}]{Sheridan2019}%
  \BibitemOpen
  \bibfield  {author} {\bibinfo {author} {\bibfnamefont {T.E.}\ \bibnamefont
  {Sheridan}}, \bibinfo {author} {\bibfnamefont {N.R.}\ \bibnamefont {Weiner}},
  \ and\ \bibinfo {author} {\bibfnamefont {T.E.}\ \bibnamefont {Steinberger}},\
  }\bibfield  {title} {\enquote {\bibinfo {title} {Dust and plasma properties
  measured using two confined particles},}\ }\href
  {http://dx.doi.org/10.1017/S002237781600060X} {\bibfield  {journal} {\bibinfo
   {journal} {J. Plasma Phys. (UK)}\ }\textbf {\bibinfo {volume} {82}},\
  \bibinfo {pages} {615820304 (10 pp.) --} (\bibinfo {year}
  {2016})}\BibitemShut {NoStop}%
\bibitem [{\citenamefont {Joshi}\ \emph {et~al.}(2023)\citenamefont {Joshi},
  \citenamefont {Pustylnik}, \citenamefont {Thoma}, \citenamefont {Thomas},\
  and\ \citenamefont {Schwabe}}]{Joshi2023}%
  \BibitemOpen
  \bibfield  {author} {\bibinfo {author} {\bibfnamefont {E.}~\bibnamefont
  {Joshi}}, \bibinfo {author} {\bibfnamefont {M.Y.}\ \bibnamefont {Pustylnik}},
  \bibinfo {author} {\bibfnamefont {M.H.}\ \bibnamefont {Thoma}}, \bibinfo
  {author} {\bibfnamefont {H.M.}\ \bibnamefont {Thomas}}, \ and\ \bibinfo
  {author} {\bibfnamefont {M.}~\bibnamefont {Schwabe}},\ }\bibfield  {title}
  {\enquote {\bibinfo {title} {Recrystallization in string-fluid complex
  plasmas},}\ }\href {http://dx.doi.org/10.1103/PhysRevResearch.5.L012030}
  {\bibfield  {journal} {\bibinfo  {journal} {Phys. Rev. Res. (USA)}\ ,\
  \bibinfo {pages} {L012030 (6 pp.) --}} (\bibinfo {year} {2023})}\BibitemShut
  {NoStop}%
\bibitem [{\citenamefont {Sargsyan}\ \emph {et~al.}(2010)\citenamefont
  {Sargsyan}, \citenamefont {Sarkisyan}, \citenamefont {Krohn}, \citenamefont
  {Keaveney},\ and\ \citenamefont {Adams}}]{Sargsyan2010}%
  \BibitemOpen
  \bibfield  {author} {\bibinfo {author} {\bibfnamefont {Armen}\ \bibnamefont
  {Sargsyan}}, \bibinfo {author} {\bibfnamefont {David}\ \bibnamefont
  {Sarkisyan}}, \bibinfo {author} {\bibfnamefont {Ulrich}\ \bibnamefont
  {Krohn}}, \bibinfo {author} {\bibfnamefont {James}\ \bibnamefont {Keaveney}},
  \ and\ \bibinfo {author} {\bibfnamefont {Charles}\ \bibnamefont {Adams}},\
  }\bibfield  {title} {\enquote {\bibinfo {title} {Effect of buffer gas on an
  electromagnetically induced transparency in a ladder system using thermal
  rubidium vapor},}\ }\href {\doibase
  10.1103/PHYSREVA.82.045806/FIGURES/4/MEDIUM} {\bibfield  {journal} {\bibinfo
  {journal} {Physical Review A - Atomic, Molecular, and Optical Physics}\
  }\textbf {\bibinfo {volume} {82}},\ \bibinfo {pages} {045806} (\bibinfo
  {year} {2010})}\BibitemShut {NoStop}%
\bibitem [{\citenamefont {Budker}\ \emph {et~al.}(2002)\citenamefont {Budker},
  \citenamefont {Gawlik}, \citenamefont {Kimball}, \citenamefont {Rochester},
  \citenamefont {Yashchuk},\ and\ \citenamefont {Weis}}]{Budker2002}%
  \BibitemOpen
  \bibfield  {author} {\bibinfo {author} {\bibfnamefont {D.}~\bibnamefont
  {Budker}}, \bibinfo {author} {\bibfnamefont {W.}~\bibnamefont {Gawlik}},
  \bibinfo {author} {\bibfnamefont {D.F.}\ \bibnamefont {Kimball}}, \bibinfo
  {author} {\bibfnamefont {S.M.}\ \bibnamefont {Rochester}}, \bibinfo {author}
  {\bibfnamefont {V.V.}\ \bibnamefont {Yashchuk}}, \ and\ \bibinfo {author}
  {\bibfnamefont {A.}~\bibnamefont {Weis}},\ }\bibfield  {title} {\enquote
  {\bibinfo {title} {Resonant nonlinear magneto-optical effects in atoms},}\
  }\href {http://dx.doi.org/10.1103/RevModPhys.74.1153} {\bibfield  {journal}
  {\bibinfo  {journal} {Rev. Mod. Phys. (USA)}\ }\textbf {\bibinfo {volume}
  {74}},\ \bibinfo {pages} {1153 -- 201} (\bibinfo {year} {2002})}\BibitemShut
  {NoStop}%
\bibitem [{\citenamefont {Allred}\ \emph {et~al.}(2002)\citenamefont {Allred},
  \citenamefont {Lyman}, \citenamefont {Kornack},\ and\ \citenamefont
  {Romalis}}]{Allred2002}%
  \BibitemOpen
  \bibfield  {author} {\bibinfo {author} {\bibfnamefont {J.~C.}\ \bibnamefont
  {Allred}}, \bibinfo {author} {\bibfnamefont {R.~N.}\ \bibnamefont {Lyman}},
  \bibinfo {author} {\bibfnamefont {T.~W.}\ \bibnamefont {Kornack}}, \ and\
  \bibinfo {author} {\bibfnamefont {M.~V.}\ \bibnamefont {Romalis}},\
  }\bibfield  {title} {\enquote {\bibinfo {title} {High-sensitivity atomic
  magnetometer unaffected by spin-exchange relaxation},}\ }\href {\doibase
  10.1103/PhysRevLett.89.130801} {\bibfield  {journal} {\bibinfo  {journal}
  {Phys. Rev. Lett.}\ }\textbf {\bibinfo {volume} {89}},\ \bibinfo {pages}
  {130801} (\bibinfo {year} {2002})}\BibitemShut {NoStop}%
\bibitem [{\citenamefont {Kornack}\ \emph {et~al.}(2005)\citenamefont
  {Kornack}, \citenamefont {Ghosh},\ and\ \citenamefont
  {Romalis}}]{Kornack2005}%
  \BibitemOpen
  \bibfield  {author} {\bibinfo {author} {\bibfnamefont {T.~W.}\ \bibnamefont
  {Kornack}}, \bibinfo {author} {\bibfnamefont {R.~K.}\ \bibnamefont {Ghosh}},
  \ and\ \bibinfo {author} {\bibfnamefont {M.~V.}\ \bibnamefont {Romalis}},\
  }\bibfield  {title} {\enquote {\bibinfo {title} {Nuclear spin gyroscope based
  on an atomic comagnetometer},}\ }\href {\doibase
  10.1103/PhysRevLett.95.230801} {\bibfield  {journal} {\bibinfo  {journal}
  {Phys. Rev. Lett.}\ }\textbf {\bibinfo {volume} {95}},\ \bibinfo {pages}
  {230801} (\bibinfo {year} {2005})}\BibitemShut {NoStop}%
\bibitem [{\citenamefont {Alekseev}\ and\ \citenamefont
  {Sobel’man}(1966)}]{alekseev1966spectroscopic}%
  \BibitemOpen
  \bibfield  {author} {\bibinfo {author} {\bibfnamefont {V~A}\ \bibnamefont
  {Alekseev}}\ and\ \bibinfo {author} {\bibfnamefont {I~I}\ \bibnamefont
  {Sobel’man}},\ }\bibfield  {title} {\enquote {\bibinfo {title} {A
  spectroscopic method for the investigation of elastic scattering of slow
  electrons},}\ }\href@noop {} {\bibfield  {journal} {\bibinfo  {journal} {Sov.
  Phys. JETP}\ }\textbf {\bibinfo {volume} {22}},\ \bibinfo {pages} {882}
  (\bibinfo {year} {1966})}\BibitemShut {NoStop}%
\bibitem [{\citenamefont {Omont}(1977)}]{Omont1977}%
  \BibitemOpen
  \bibfield  {author} {\bibinfo {author} {\bibfnamefont {A.}~\bibnamefont
  {Omont}},\ }\bibfield  {title} {\enquote {\bibinfo {title} {{on the Theory of
  Collisions of Atoms in Rydberg States With Neutral Particles.}}}\ }\href
  {\doibase 10.1051/jphys:0197700380110134300} {\bibfield  {journal} {\bibinfo
  {journal} {J Phys (Paris)}\ }\textbf {\bibinfo {volume} {38}},\ \bibinfo
  {pages} {1345--1359} (\bibinfo {year} {1977})}\BibitemShut {NoStop}%
\bibitem [{\citenamefont {Asaf}\ \emph {et~al.}(1993)\citenamefont {Asaf},
  \citenamefont {Rupnik}, \citenamefont {Reisfeld},\ and\ \citenamefont
  {McGlynn}}]{Asaf1993}%
  \BibitemOpen
  \bibfield  {author} {\bibinfo {author} {\bibfnamefont {U.}~\bibnamefont
  {Asaf}}, \bibinfo {author} {\bibfnamefont {K.}~\bibnamefont {Rupnik}},
  \bibinfo {author} {\bibfnamefont {G.}~\bibnamefont {Reisfeld}}, \ and\
  \bibinfo {author} {\bibfnamefont {S.~P.}\ \bibnamefont {McGlynn}},\
  }\bibfield  {title} {\enquote {\bibinfo {title} {{Pressure shifts and
  electron scattering lengths in atomic and molecular gases}},}\ }\href
  {\doibase 10.1063/1.465219} {\bibfield  {journal} {\bibinfo  {journal} {The
  Journal of Chemical Physics}\ }\textbf {\bibinfo {volume} {99}},\ \bibinfo
  {pages} {2560--2566} (\bibinfo {year} {1993})}\BibitemShut {NoStop}%
\bibitem [{\citenamefont {Fermi}(1934)}]{fermi1934sopra}%
  \BibitemOpen
  \bibfield  {author} {\bibinfo {author} {\bibfnamefont {Enrico}\ \bibnamefont
  {Fermi}},\ }\bibfield  {title} {\enquote {\bibinfo {title} {Sopra lo
  spostamento per pressione delle righe elevate delle serie spettrali},}\
  }\href@noop {} {\bibfield  {journal} {\bibinfo  {journal} {Il Nuovo Cimento
  (1924-1942)}\ }\textbf {\bibinfo {volume} {11}},\ \bibinfo {pages} {157--166}
  (\bibinfo {year} {1934})}\BibitemShut {NoStop}%
\bibitem [{\citenamefont {Brillet}\ and\ \citenamefont
  {Gallagher}(1980)}]{Brillet1980}%
  \BibitemOpen
  \bibfield  {author} {\bibinfo {author} {\bibfnamefont {Wan-\"U~L.}\
  \bibnamefont {Brillet}}\ and\ \bibinfo {author} {\bibfnamefont
  {A.}~\bibnamefont {Gallagher}},\ }\bibfield  {title} {\enquote {\bibinfo
  {title} {Inert-gas collisional broadening and shifts of rb rydberg states},}\
  }\href {\doibase 10.1103/PhysRevA.22.1012} {\bibfield  {journal} {\bibinfo
  {journal} {Phys. Rev. A}\ }\textbf {\bibinfo {volume} {22}},\ \bibinfo
  {pages} {1012--1017} (\bibinfo {year} {1980})}\BibitemShut {NoStop}%
\bibitem [{\citenamefont {Weber}\ and\ \citenamefont
  {Niemax}(1982)}]{weber1982impact}%
  \BibitemOpen
  \bibfield  {author} {\bibinfo {author} {\bibfnamefont {K~H}\ \bibnamefont
  {Weber}}\ and\ \bibinfo {author} {\bibfnamefont {K}~\bibnamefont {Niemax}},\
  }\bibfield  {title} {\enquote {\bibinfo {title} {Impact broadening and shift
  of rb ns and nd levels by noble gases},}\ }\href@noop {} {\bibfield
  {journal} {\bibinfo  {journal} {Zeitschrift f{\"u}r Physik A Atoms and
  Nuclei}\ }\textbf {\bibinfo {volume} {307}},\ \bibinfo {pages} {13--24}
  (\bibinfo {year} {1982})}\BibitemShut {NoStop}%
\bibitem [{\citenamefont {Bruce}\ \emph {et~al.}(1982)\citenamefont {Bruce},
  \citenamefont {Mirza},\ and\ \citenamefont {Duley}}]{bruce1982collision}%
  \BibitemOpen
  \bibfield  {author} {\bibinfo {author} {\bibfnamefont {D~M}\ \bibnamefont
  {Bruce}}, \bibinfo {author} {\bibfnamefont {M~Y}\ \bibnamefont {Mirza}}, \
  and\ \bibinfo {author} {\bibfnamefont {W~W}\ \bibnamefont {Duley}},\
  }\bibfield  {title} {\enquote {\bibinfo {title} {Collision broadening and
  shift of the s and d rydberg levels of rubidium by he and ar},}\ }\href@noop
  {} {\bibfield  {journal} {\bibinfo  {journal} {Optics Communications}\
  }\textbf {\bibinfo {volume} {40}},\ \bibinfo {pages} {347--352} (\bibinfo
  {year} {1982})}\BibitemShut {NoStop}%
\bibitem [{\citenamefont {Thompson}\ \emph {et~al.}(1987)\citenamefont
  {Thompson}, \citenamefont {Kammermayer}, \citenamefont {Stoicheff},\ and\
  \citenamefont {Weinberger}}]{Thompson1987}%
  \BibitemOpen
  \bibfield  {author} {\bibinfo {author} {\bibfnamefont {D.~C.}\ \bibnamefont
  {Thompson}}, \bibinfo {author} {\bibfnamefont {E.}~\bibnamefont
  {Kammermayer}}, \bibinfo {author} {\bibfnamefont {B.~P.}\ \bibnamefont
  {Stoicheff}}, \ and\ \bibinfo {author} {\bibfnamefont {E.}~\bibnamefont
  {Weinberger}},\ }\bibfield  {title} {\enquote {\bibinfo {title} {Pressure
  shifts and broadenings of rb rydberg states by ne, kr, and
  ${\mathrm{h}}_{2}$},}\ }\href {\doibase 10.1103/PhysRevA.36.2134} {\bibfield
  {journal} {\bibinfo  {journal} {Phys. Rev. A}\ }\textbf {\bibinfo {volume}
  {36}},\ \bibinfo {pages} {2134--2141} (\bibinfo {year} {1987})}\BibitemShut
  {NoStop}%
\bibitem [{\citenamefont {Ottinger}\ \emph {et~al.}(1975)\citenamefont
  {Ottinger}, \citenamefont {Scheps}, \citenamefont {York},\ and\ \citenamefont
  {Gallagher}}]{Ottinger1975}%
  \BibitemOpen
  \bibfield  {author} {\bibinfo {author} {\bibfnamefont {Ch}~\bibnamefont
  {Ottinger}}, \bibinfo {author} {\bibfnamefont {Richard}\ \bibnamefont
  {Scheps}}, \bibinfo {author} {\bibfnamefont {G.~W.}\ \bibnamefont {York}}, \
  and\ \bibinfo {author} {\bibfnamefont {Alan}\ \bibnamefont {Gallagher}},\
  }\bibfield  {title} {\enquote {\bibinfo {title} {Broadening of the rb
  resonance lines by the noble gases},}\ }\href {\doibase
  10.1103/PhysRevA.11.1815} {\bibfield  {journal} {\bibinfo  {journal}
  {Physical Review A}\ }\textbf {\bibinfo {volume} {11}},\ \bibinfo {pages}
  {1815} (\bibinfo {year} {1975})}\BibitemShut {NoStop}%
\bibitem [{\citenamefont {Ockenfels}\ \emph {et~al.}(2022)\citenamefont
  {Ockenfels}, \citenamefont {Roje}, \citenamefont {vom H{\"o}vel},
  \citenamefont {Vewinger},\ and\ \citenamefont
  {Weitz}}]{ockenfels2022spectroscopy}%
  \BibitemOpen
  \bibfield  {author} {\bibinfo {author} {\bibfnamefont {Till}\ \bibnamefont
  {Ockenfels}}, \bibinfo {author} {\bibfnamefont {Pa{\v{s}}ko}\ \bibnamefont
  {Roje}}, \bibinfo {author} {\bibfnamefont {Thilo}\ \bibnamefont {vom
  H{\"o}vel}}, \bibinfo {author} {\bibfnamefont {Frank}\ \bibnamefont
  {Vewinger}}, \ and\ \bibinfo {author} {\bibfnamefont {Martin}\ \bibnamefont
  {Weitz}},\ }\bibfield  {title} {\enquote {\bibinfo {title} {Spectroscopy of
  high-pressure rubidium--noble-gas mixtures},}\ }\href@noop {} {\bibfield
  {journal} {\bibinfo  {journal} {Physical Review A}\ }\textbf {\bibinfo
  {volume} {106}},\ \bibinfo {pages} {012815} (\bibinfo {year}
  {2022})}\BibitemShut {NoStop}%
\bibitem [{\citenamefont {Henry}\ and\ \citenamefont
  {Herman}(2002)}]{Henry2002}%
  \BibitemOpen
  \bibfield  {author} {\bibinfo {author} {\bibfnamefont {Megan~E.}\
  \bibnamefont {Henry}}\ and\ \bibinfo {author} {\bibfnamefont {Roger~M.}\
  \bibnamefont {Herman}},\ }\bibfield  {title} {\enquote {\bibinfo {title}
  {{Collisional broadening of Rydberg atom transitions by rare gas
  perturbers}},}\ }\href {\doibase 10.1088/0953-4075/35/2/313} {\bibfield
  {journal} {\bibinfo  {journal} {Journal of Physics B: Atomic, Molecular and
  Optical Physics}\ }\textbf {\bibinfo {volume} {35}},\ \bibinfo {pages} {373}
  (\bibinfo {year} {2002})}\BibitemShut {NoStop}%
\bibitem [{\citenamefont {Hoffmann}\ and\ \citenamefont
  {Skarsgard}(1969)}]{Hoffmann1969}%
  \BibitemOpen
  \bibfield  {author} {\bibinfo {author} {\bibfnamefont {C.~R.}\ \bibnamefont
  {Hoffmann}}\ and\ \bibinfo {author} {\bibfnamefont {H.~M.}\ \bibnamefont
  {Skarsgard}},\ }\bibfield  {title} {\enquote {\bibinfo {title}
  {Momentum-transfer cross sections and conductivity ratios for low-energy
  electrons in he, ne, kr, and xe},}\ }\href {\doibase 10.1103/PhysRev.178.168}
  {\bibfield  {journal} {\bibinfo  {journal} {Phys. Rev.}\ }\textbf {\bibinfo
  {volume} {178}},\ \bibinfo {pages} {168--175} (\bibinfo {year}
  {1969})}\BibitemShut {NoStop}%
\bibitem [{\citenamefont {Fedus}(2014)}]{Fedus2014}%
  \BibitemOpen
  \bibfield  {author} {\bibinfo {author} {\bibfnamefont {Kamil}\ \bibnamefont
  {Fedus}},\ }\bibfield  {title} {\enquote {\bibinfo {title} {Electron
  scattering from neon via effective range theory},}\ }\href {\doibase
  10.1007/s13538-014-0265-z} {\bibfield  {journal} {\bibinfo  {journal}
  {Brazilian Journal of Physics}\ }\textbf {\bibinfo {volume} {44}},\ \bibinfo
  {pages} {622--628} (\bibinfo {year} {2014})}\BibitemShut {NoStop}%
\bibitem [{\citenamefont {Cheng}\ \emph {et~al.}(2014)\citenamefont {Cheng},
  \citenamefont {Tang}, \citenamefont {Mitroy},\ and\ \citenamefont
  {Safronova}}]{Cheng2014}%
  \BibitemOpen
  \bibfield  {author} {\bibinfo {author} {\bibfnamefont {Yongjun}\ \bibnamefont
  {Cheng}}, \bibinfo {author} {\bibfnamefont {Li~Yan}\ \bibnamefont {Tang}},
  \bibinfo {author} {\bibfnamefont {J.}~\bibnamefont {Mitroy}}, \ and\ \bibinfo
  {author} {\bibfnamefont {M.~S.}\ \bibnamefont {Safronova}},\ }\bibfield
  {title} {\enquote {\bibinfo {title} {All-order relativistic many-body theory
  of low-energy electron-atom scattering},}\ }\href {\doibase
  10.1103/PhysRevA.89.012701} {\bibfield  {journal} {\bibinfo  {journal} {Phys.
  Rev. A}\ }\textbf {\bibinfo {volume} {89}},\ \bibinfo {pages} {012701}
  (\bibinfo {year} {2014})}\BibitemShut {NoStop}%
\bibitem [{\citenamefont {Steck}(2021)}]{Steck}%
  \BibitemOpen
  \bibfield  {author} {\bibinfo {author} {\bibfnamefont {Daniel~A.}\
  \bibnamefont {Steck}},\ }\href {https://steck.us/alkalidata/} {\enquote
  {\bibinfo {title} {Rubidium 85 d line data},}\ } (\bibinfo {year}
  {2021})\BibitemShut {NoStop}%
\end{thebibliography}%

\end{document}